\documentclass[prd, showpacs, twocolumn,superscriptaddress]{revtex4}

\usepackage{graphicx}   
\begin{document}
\preprint{version 4.2} 

\title{The first search for extremely-high energy cosmogenic neutrinos\\ 
with the IceCube Neutrino Observatory}

\affiliation{III. Physikalisches Institut, RWTH Aachen University, D-52056 Aachen, Germany}
\affiliation{Dept.~of Physics and Astronomy, University of Alabama, Tuscaloosa, AL 35487, USA}
\affiliation{Dept.~of Physics and Astronomy, University of Alaska Anchorage, 3211 Providence Dr., Anchorage, AK 99508, USA}
\affiliation{CTSPS, Clark-Atlanta University, Atlanta, GA 30314, USA}
\affiliation{School of Physics and Center for Relativistic Astrophysics, Georgia Institute of Technology, Atlanta, GA 30332, USA}
\affiliation{Dept.~of Physics, Southern University, Baton Rouge, LA 70813, USA}
\affiliation{Dept.~of Physics, University of California, Berkeley, CA 94720, USA}
\affiliation{Lawrence Berkeley National Laboratory, Berkeley, CA 94720, USA}
\affiliation{Institut f\"ur Physik, Humboldt-Universit\"at zu Berlin, D-12489 Berlin, Germany}
\affiliation{Fakult\"at f\"ur Physik \& Astronomie, Ruhr-Universit\"at Bochum, D-44780 Bochum, Germany}
\affiliation{Physikalisches Institut, Universit\"at Bonn, Nussallee 12, D-53115 Bonn, Germany}
\affiliation{Dept.~of Physics, University of the West Indies, Cave Hill Campus, Bridgetown BB11000, Barbados}
\affiliation{Universit\'e Libre de Bruxelles, Science Faculty CP230, B-1050 Brussels, Belgium}
\affiliation{Vrije Universiteit Brussel, Dienst ELEM, B-1050 Brussels, Belgium}
\affiliation{Dept.~of Physics, Chiba University, Chiba 263-8522, Japan}
\affiliation{Dept.~of Physics and Astronomy, University of Canterbury, Private Bag 4800, Christchurch, New Zealand}
\affiliation{Dept.~of Physics, University of Maryland, College Park, MD 20742, USA}
\affiliation{Dept.~of Physics and Center for Cosmology and Astro-Particle Physics, Ohio State University, Columbus, OH 43210, USA}
\affiliation{Dept.~of Astronomy, Ohio State University, Columbus, OH 43210, USA}
\affiliation{Dept.~of Physics, TU Dortmund University, D-44221 Dortmund, Germany}
\affiliation{Dept.~of Physics, University of Alberta, Edmonton, Alberta, Canada T6G 2G7}
\affiliation{Dept.~of Subatomic and Radiation Physics, University of Gent, B-9000 Gent, Belgium}
\affiliation{Max-Planck-Institut f\"ur Kernphysik, D-69177 Heidelberg, Germany}
\affiliation{Dept.~of Physics and Astronomy, University of California, Irvine, CA 92697, USA}
\affiliation{Laboratory for High Energy Physics, \'Ecole Polytechnique F\'ed\'erale, CH-1015 Lausanne, Switzerland}
\affiliation{Dept.~of Physics and Astronomy, University of Kansas, Lawrence, KS 66045, USA}
\affiliation{Dept.~of Astronomy, University of Wisconsin, Madison, WI 53706, USA}
\affiliation{Dept.~of Physics, University of Wisconsin, Madison, WI 53706, USA}
\affiliation{Institute of Physics, University of Mainz, Staudinger Weg 7, D-55099 Mainz, Germany}
\affiliation{Universit\'e de Mons, 7000 Mons, Belgium}
\affiliation{Bartol Research Institute and Department of Physics and Astronomy, University of Delaware, Newark, DE 19716, USA}
\affiliation{Dept.~of Physics, University of Oxford, 1 Keble Road, Oxford OX1 3NP, UK}
\affiliation{Dept.~of Physics, University of Wisconsin, River Falls, WI 54022, USA}
\affiliation{Oskar Klein Centre and Dept.~of Physics, Stockholm University, SE-10691 Stockholm, Sweden}
\affiliation{Dept.~of Astronomy and Astrophysics, Pennsylvania State University, University Park, PA 16802, USA}
\affiliation{Dept.~of Physics, Pennsylvania State University, University Park, PA 16802, USA}
\affiliation{Dept.~of Physics and Astronomy, Uppsala University, Box 516, S-75120 Uppsala, Sweden}
\affiliation{Dept.~of Physics and Astronomy, Utrecht University/SRON, NL-3584 CC Utrecht, The Netherlands}
\affiliation{Dept.~of Physics, University of Wuppertal, D-42119 Wuppertal, Germany}
\affiliation{DESY, D-15735 Zeuthen, Germany}

\author{R.~Abbasi}
\affiliation{Dept.~of Physics, University of Wisconsin, Madison, WI 53706, USA}
\author{Y.~Abdou}
\affiliation{Dept.~of Subatomic and Radiation Physics, University of Gent, B-9000 Gent, Belgium}
\author{T.~Abu-Zayyad}
\affiliation{Dept.~of Physics, University of Wisconsin, River Falls, WI 54022, USA}
\author{J.~Adams}
\affiliation{Dept.~of Physics and Astronomy, University of Canterbury, Private Bag 4800, Christchurch, New Zealand}
\author{J.~A.~Aguilar}
\affiliation{Dept.~of Physics, University of Wisconsin, Madison, WI 53706, USA}
\author{M.~Ahlers}
\affiliation{Dept.~of Physics, University of Oxford, 1 Keble Road, Oxford OX1 3NP, UK}
\author{K.~Andeen}
\affiliation{Dept.~of Physics, University of Wisconsin, Madison, WI 53706, USA}
\author{J.~Auffenberg}
\affiliation{Dept.~of Physics, University of Wuppertal, D-42119 Wuppertal, Germany}
\author{X.~Bai}
\affiliation{Bartol Research Institute and Department of Physics and Astronomy, University of Delaware, Newark, DE 19716, USA}
\author{M.~Baker}
\affiliation{Dept.~of Physics, University of Wisconsin, Madison, WI 53706, USA}
% ...

\collaboration{IceCube Collaboration}
\noaffiliation

\date{\today}

\begin{abstract}
 %
 %
 %
 We report on the results of the search for extremely-high
 energy (EHE) neutrinos with energies
 above $10^7$ GeV obtained with the partially ($\sim$30\%)
 constructed IceCube in 2007.
 From the absence of signal events in the sample of 242.1 days of effective
 livetime, we derive a 90\% C.L. model independent differential upper limit
 based on the number of signal events per energy decade 
 at $E^2 \phi_{\nu_e+\nu_\mu+\nu_\tau}\simeq 1.4 \times 10^{-6}$
 GeV cm$^{-2}$ sec$^{-1}$ sr$^{-1}$ for neutrinos in the energy range from $3\times10^7$
 to $3\times10^9$ GeV.
\end{abstract}
 

\maketitle

%
%
\section{\label{sec:intro} Introduction}
Detection of extremely-high energy (EHE) neutrinos with
energies greater than $10^7$~GeV may shed light on the long standing
puzzle of the origin of EHE cosmic-rays~\cite{cronin99, nagano01}.
Several observational results have indicated that these EHE cosmic 
rays (EHECRs) are of extragalactic origin~\cite{auger07}. 
Further elucidation of their production mechanism by EHECR observation
is, however, limited because the collisions of EHECR with the cosmic 
microwave background photons, known as the Greisen-Zatsepin-Kuzmin (GZK) 
mechanism~\cite{GZK}, prevent EHECRs from propagating over cosmological 
distances without losing a sizable fraction of their energy. 
On the other hand, cosmogenic neutrinos~\cite{berezinsky69} produced by 
the GZK mechanism via photo-produced $\pi$ meson decay as
$\pi^{\pm}\to \mu^{\pm}\nu_\mu\to e^{\pm}\nu_e\nu_\mu$ carry information 
on the EHECR source evolution and the maximum energy of EHECRs at their 
production sites~\cite{yoshida93}.

Detection of these EHE neutrinos
is an experimental challenge because 
the very low EHE neutrino fluxes require 
a very large detector.  
The large size of the IceCube neutrino observatory~\cite{IceCubePhysics}, 
currently under construction at the geographic South Pole, will make it
more effective than previous experiments in the search for
these neutrinos~\cite{halzen06,yoshida04}.
Interactions of $\nu_\mu$, $\nu_e$, and $\nu_\tau$
and their antiparticles are observed through the Cherenkov
radiation emitted by secondary particles. 
In the following, we do not distinguish between 
$\nu$ and $\overline\nu$;
the simulations and sensitivity calculations assume an equal 
mixture of particles and antiparticles.

In this paper we will describe the first results of a search for
signatures of cosmogenic neutrinos in the 2007 data acquired by the partially
constructed IceCube neutrino observatory.
This analysis selects events which produce a large amount of light 
in the detector. Based on simple criteria, 
such as the total number of observed Cherenkov photons 
and the results of reconstruction algorithms, 
it selects candidate neutrino events.   
Although $\nu_\mu$, $\nu_e$, and $\nu_\tau$ interactions look very different 
in IceCube, the selection criteria are sensitive to all three flavors.  


%
%
\section{\label{sec:detector} The IceCube Detector}
IceCube is a cubic-kilometer, 
high-energy cosmic neutrino telescope which is currently
under construction.
%
It uses the 2800~m thick glacial ice as a Cherenkov medium. 
Cherenkov photons emitted 
by relativistic charged particles,
notably muons,
electrons, and taus produced in charged current interactions 
and their secondaries, are detected 
by an array of photon sensors, known as
Digital Optical Modules (DOMs) \cite{DOM}.
The DOMs deep below the ice surface are deployed along electrical cable
bundles that carry power and communication between the DOMs and surface
electronics. The cable assemblies, often called strings, are 
lowered into holes drilled to a depth of 2450~m. 
The DOMs, spaced at intervals of 17~m,  occupy the 
bottom 1000~meters of each string. The strings are
arranged in a hexagonal lattice pattern with a spacing of approximately
125~m. 
DOMs are also frozen into tanks located at the surface near the top of
each hole. The tanks constitute an air shower array called IceTop
\cite{icetop}. 

The DOMs enclose a down-looking 25~cm photomultiplier tube (PMT) \cite{PMT}
with data acquisition and calibration electronics, 
light emitting diodes for calibration,
and also data compression, communications, and control hardware~\cite{DOM}
in a 35~cm diameter pressure sphere.  
Almost all of the PMTs are run at a gain of $10^{7}$; 
PMT saturation effects become important at signal levels of about 
5000 photoelectrons in 
a single DOM in 50 ns.  
When the DOM detects a photoelectron, 
it initiates an acquisition cycle, recording the PMT output 
with two waveform digitizer systems.  
The first system samples every 3.3 ns for 400 ns, with 14 bits of dynamic range. 
The second system samples 
every 25 ns for 6.4 $\mu$s, with 10 bits of dynamic range.  
The data acquisition system is designed
such that the first system is sensitive to a bright
photon source at close distance and the second system 
captures signal induced by photons emitted at large distance. 
This analysis uses the total number of photoelectrons detected by the PMTs
as a measure of the event energy.  
For each DOM, the charge used is the one from 
whichever system recorded a larger number of photoelectrons.
Because of the significant DOM-to-DOM differences in saturation behavior, 
the current analysis does not attempt to correct for PMT saturation. 
So, the signals from brightly illuminated DOMs are naturally truncated.

%
%
\section{\label{sec:simulation-dataset} 
 Data and Simulation}

This analysis uses data collected from May 2007 through April 2008, 
when IceCube consisted of 22 in-ice strings 
(\mbox{IC-22}; 1320 DOMs) and 52 IceTop tanks.
In order to greatly reduce random noise from radioactivity, in IC22
the DOMs only recorded signal waveforms when a local coincidence (LC)
condition was satisfied, {\it i.e.} when an adjoining 
or next-to-nearest neighbor DOM was triggered
within $\pm$1~$\mu$s.
In 2007, the trigger selected time periods when 8 or more 
DOMs recorded LC signals within  5 $\mu$s; when this happened, all hits within a
20$\mu$s window were stored as an event.  
The average trigger rate was about 550 Hz. 
The high-multiplicity event sample used in this analysis 
imposes an additional condition requiring
NDOM $\ge 80$,  where NDOM is the number of hit DOMs in an event.    
The average high-multiplicity event rate was approximately 1.5 Hz with a
seasonal variation of 17\%.
A total of $3.2 \times 10^{7}$ events were tagged as high-multiplicity during the
effective livetime of 242.1 days (excluding the periods of unstable operation).

The high multiplicity cut reduces the data by a factor of $\sim 3\times 10^{-3}$
while preserving approximately 70\% of the GZK neutrinos
with projected trajectories that pass 
within 880 m of the center of IceCube.
Here, and below, the GZK signal rates are based on 
the GZK spectra and flux calculated 
by Ref.~\cite{yoshida93} assuming 
an all-proton composition with a moderately strong source evolution, 
$(z+1)^m$ with $m=4$ extending to $z=4.0$. 
Neutrino oscillations modify the neutrino flavor ratio 
over the cosmological distances they travel
and the fluxes at the Earth were calculated
as in Ref.~\cite{jones04}. Note that
the flavor ratio $\nu_e: \nu_\mu:\nu_\tau$ of cosmogenic neutrinos
at the Earth is different from 1:1:1
as primary energy spectra of $\nu_e$ and $\nu_{\mu}$ 
produced by the GZK mechanism are different
because of a significant contribution of $\nu_e$ from neutron decay.
This enhancement was included in the GZK neutrino flux calculations
used here.

EHE neutrinos were simulated with the JULIeT package~\cite{yoshida04}
to generate and propagate the neutrinos through the Earth.
All three flavors of neutrinos were simulated with energies
between $10^5$ and $10^{11}$ GeV. The resulting secondary muons and taus produced 
in the neutrino interactions are propagated through the rock and ice near the IceCube
volume, also by JULIeT. 
Hadronic and electromagnetic showers are also simulated; all of these showers are
treated as point sources, without accounting for the LPM effect.
The background muon bundles from cosmic-rays in the energy 
range $10^6$ GeV to $10^{10}$ GeV were generated using CORSIKA~\cite{corsika} 
version 6.720 with the SIBYLL 2.1 hadronic interaction model or 
with QGSJET II, without charm production~\cite{promptMu01}. 
The uncertain prompt muon component from charm decay 
may contribute to the background events~\cite{enberg08}.
The muons were propagated through the Earth using MMC~\cite{mmc}.
Departures of observed data distributions from those of 
the CORSIKA based background events prompted us to also 
develop a phenomenological background model based on fits to data.
Emission of Cherenkov photons and their propagation in the ice was
simulated by the Photonics package~\cite{photonics}. 
Measurement of the absolute number of Cherenkov
photons is important in the EHE neutrino search as it
closely relates to the energy of the high energy muons, taus or electrons produced by
EHE neutrinos. Therefore the detection efficiency of the DOMs must be
understood with good precision. The primary element, the PMT, is
calibrated in the laboratory using a nitrogen laser to measure the
photon detection efficiency~\cite{PMT}. 
This bare PMT data are used
in a simulation package which propagates
photons inside the glass sphere and the optical gel to the photo-cathode
surface. 
The DOM simulation is followed by waveform calibration and 
the trigger condition is included in the simulation chain. 
%
%
\section{Analysis}
\label{sec:analysis}

%
\subsection{\label{sec:signal} Extremely high energy event signatures and
  the initial event filter}
The event signatures from $\nu_\mu$, $\nu_e$, and $\nu_\tau$ are very different.
IceCube mainly detects cosmogenic neutrinos by the signals from the
secondary muons and taus generated in the neutrino interaction in the
rock or ice. At high energies, these particles are seen in the detector
as series of energetic cascades from radiative energy loss processes
such as pair creation, bremsstrahlung, and photonuclear interactions,
rather than minimum-ionizing tracks. The radiative energy losses are
approximately proportional to the energy of the muon or the tau, and so
is the Cherenkov light yield.  Electron neutrinos produce electromagnetic and 
hadronic showers, which are relatively compact sources of Cherenkov light. 
Muon and tau
neutrinos within IceCube will also produce a hadronic shower from the struck nucleon, 
in addition to the muon or tau secondary.

\begin{table*}
  \caption{Number of events at 
    different filter levels for 242.1 days in 2007. 
    The simulation predictions for the atmospheric muon
    background using the CORSIKA-SIBYLL package, the empirical model, and
    that for the GZK cosmogenic neutrino model are also listed for comparison.
    Errors shown here are statistical only. Refer to
 Sec.~\ref{sec:signal} for the definitions.}
  \label{tab:initial_filter}
  \centering
 \begin{tabular}{lccccc}
  \hline\hline
  Filter levels & observational  data & empirical model & CORSIKA
  (iron) & CORSIKA (proton) & signal (GZK1~\cite{yoshida93}) \\ 
  \hline
  level 1 (NDOM $\geq 80$) & $3.195 \times 10^7$ & - & (1.84$\pm$ 0.08)$\times 10^7$ & $(7.71\pm 0.45) \times 10^6$ & (886 $\pm$ 8.9)$\times 10^{-3}$ \\ 
  level 2 (NPE $> 10^4$) & 6528 & (6.82 $\pm$ 0.42)$\times 10^3$
  & (1.09 $\pm$ 0.09)$\times 10^4$ & (1.63 $\pm$ 0.17) $\times 10^3$ & (670 $\pm$ 7.5)$\times 10^{-3}$ \\ 
  \hline  \hline
 \end{tabular}
\end{table*}
%

\begin{figure}
 \includegraphics[height=2.4in, width=2.3in]{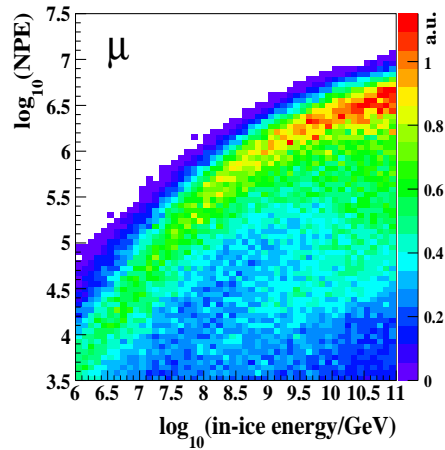}
  \caption{Event distribution from Monte Carlo simulations of single muons with
    the \mbox{IC-22} detector configurations 
    in a plane of NPE and simulated true energy.  The muon energy is given when the 
    muon is 880 meters
    from the IceCube center (\mbox{in-ice energy}). 
    The 80 DOM multiplicity cut (level 1 cut) is applied.
    The charged lepton energy distribution is assumed to follow $E^{-1}$
    in this plot for illustrative purposes.
    Only particles with trajectories intersecting within 880~{\rm m} from
    the center of IceCube array are considered in
    the plots. More distant events do not contribute to the data sample.
    }
    \label{Fig:EnergyNPE}
\end{figure}

Shown in Fig.~\ref{Fig:EnergyNPE} is the simulated distribution of the
total number of photo-electrons per event (NPE) recorded by the
the \mbox{IC-22} detector as a function of the simulated true muon energy.
A clear correlation 
between NPE and the energy of particles measured near IceCube is observed. 
The energies are sampled at a radius of 880 m from the IceCube center. 
This definition of energy is labeled "in-ice energy" 
and used throughout this paper. 
The visible departure from linearity for large NPE stems 
from the saturation of the detector during signal capture. 
%
%
Approximately 30\% of EHE signal events are due to neutrino 
interactions inside the IceCube detector volume initiating a hadronic or 
electromagnetic cascade. The correlation between NPE and incoming neutrino 
energy also holds for these events.
Electron neutrinos are detectable via this channel.

Because the energy spectrum of background atmospheric muons (both single
muons and bundles) falls steeply with energy, the GZK
neutrino flux should dominate over background in the high NPE region.
Since the through-going muons and taus induced by EHE neutrinos
enter into the IceCube volume mainly 
horizontally~\cite{halzen06,yoshida04},
the signal search criteria are
chosen to favor roughly horizontal high NPE events.

The high-multiplicity NDOM $\ge 80$ sample is dominated by
atmospheric background muons. 
The next step of the analysis selects events with NPE $>10^4$.
This reduces  the background by three orders of magnitude, leaving
6528 events, still dominated by background,
while the GZK signal reduction is $\sim$24\%. 

Table~\ref{tab:initial_filter} summarizes the number of events remaining at
each level of the initial filtering. 
In order to estimate the background in the very high energy region, 
the simulated data are compared to the experimental data in the region
$10^{4} <$ NPE $< 10^{5}$. 
The present analysis follows the blind analysis technique.
In keeping with the IceCube blindness policy, events with NPE $\geq 10^5$ were not used
for determining the background or setting cuts.
This NPE threshold was chosen so that the possible contribution 
from signal events in the studied sample was negligible.


%
%
\subsection{High-energy muon background}
\label{sec:BG}

%
\begin{figure*}
  \includegraphics[height=1.9in, width=1.7in]{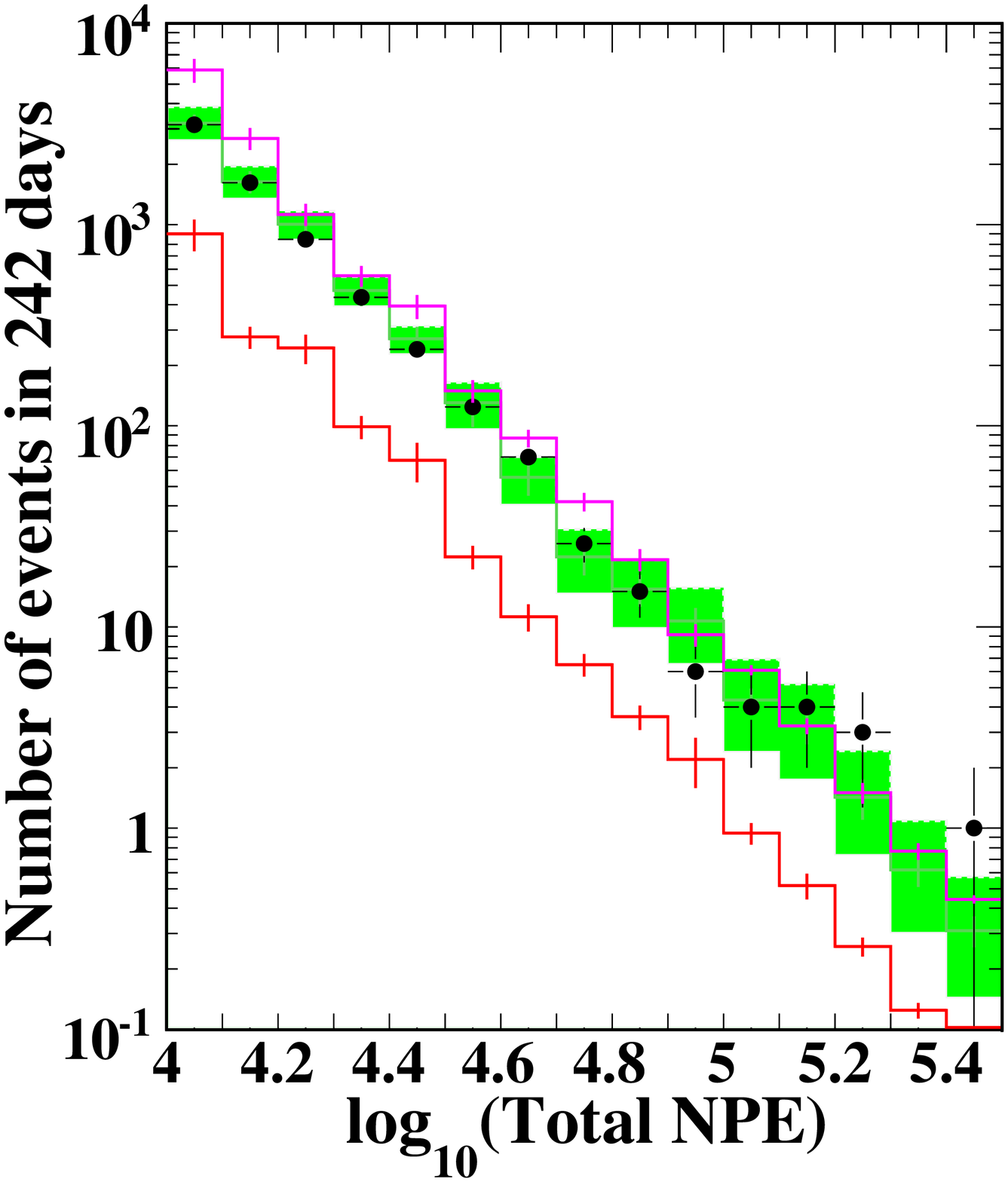}
  \includegraphics[height=1.9in, width=1.7in]{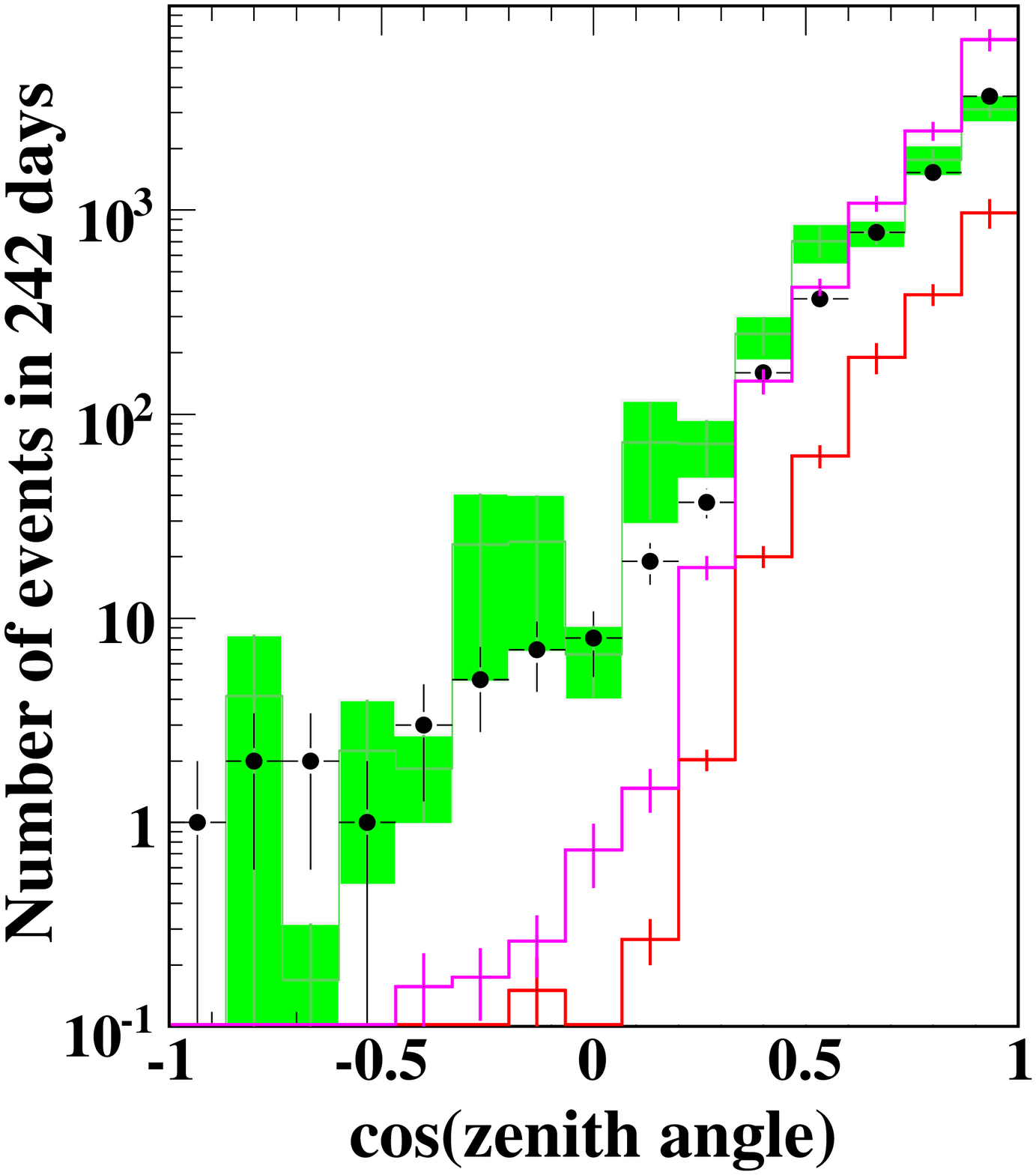}
  \includegraphics[height=1.9in, width=1.7in]{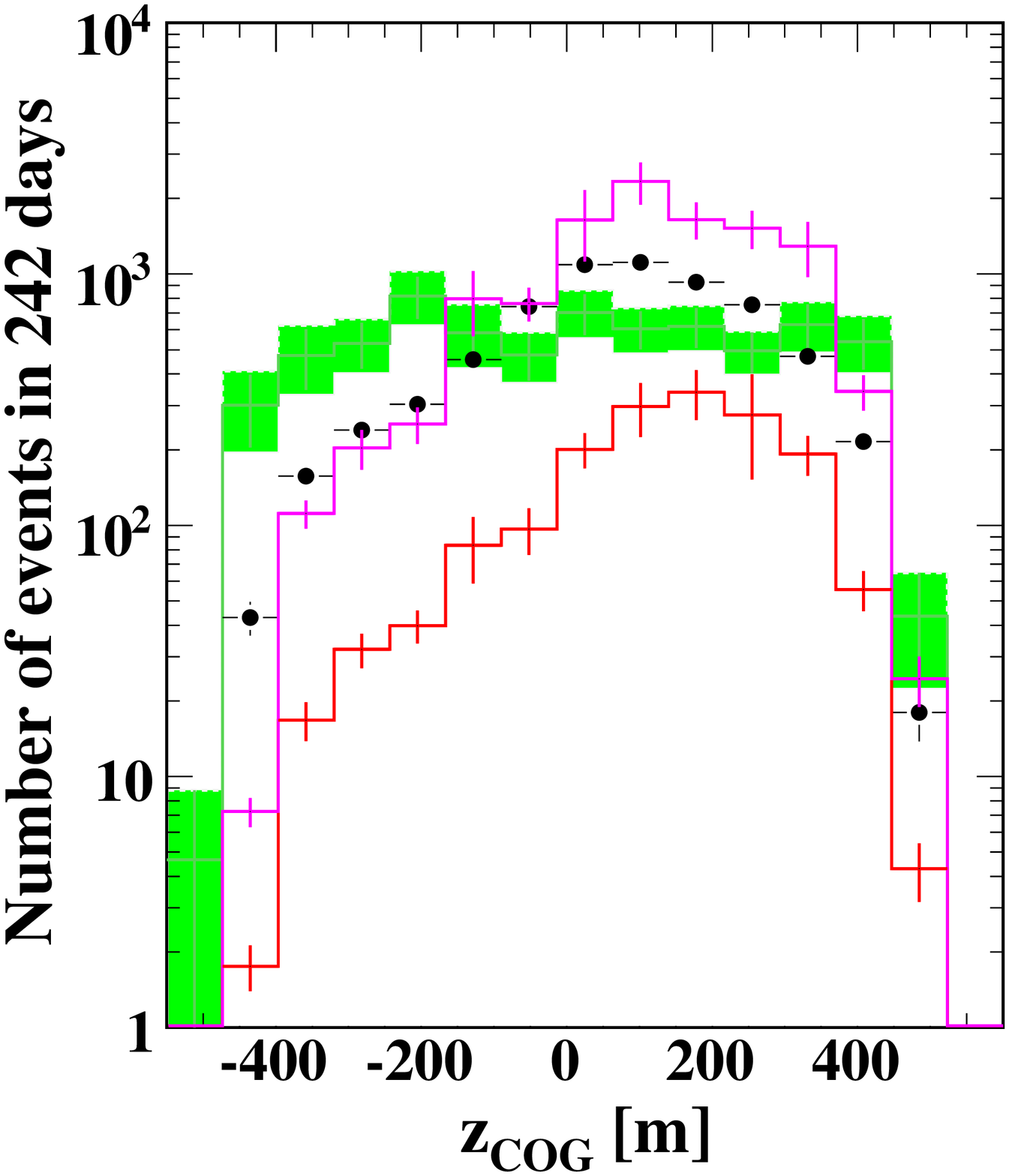}
  \caption{Event distributions for NPE, 
    cosine of reconstructed zenith angle, and the NPE-weighted 
    mean depth of event ($z_{COG}$) for observational and the background MC
    simulation data.
    The black dots represent observational data after the NPE $> 10^4$ cut, 
    red for CORSIKA proton (SIBYLL), 
    magenta for CORSIKA iron (SIBYLL). 
    Green shaded regions represent distributions obtained with the empirical model  
    with the size of shade expressing the uncertainty of the model. See text for
    the details.
 \label{BGcompCorsika}}
\end{figure*}
Bundles of muons generated in cosmic-ray air showers 
are the major background for the EHE neutrino signal search, 
because multiple muon tracks with a small geometrical 
separation resemble a single high energy muon in the IceCube detector.
The multiplicity, energy distribution, and separation distances 
for these muon bundles are not fully understood. 
Two independent Monte Carlo simulations are carried out to estimate 
the muon-bundle background in this EHE neutrino signal search. 

The first is the full cosmic-ray air shower simulation 
with light and heavy ion primaries using 
the CORSIKA (SIBYLL) package~\cite{corsika}. 
Two extreme cases of composition are used 
to address the event rate variation due to the uncertainty in
the primary cosmic-ray mass population. 
While the full air shower simulation 
includes a calculation of the production spectra 
of the multiple muons from meson decay, the simulation still introduces a 
large uncertainty because both the primary composition at relevant cosmic-ray 
energies ($> 10^7$~GeV) and the hadronic interaction model are highly uncertain. 

The second simulation uses a model relying 
on a phenomenological fit to part of the experimental high energy data. 
This empirical model approximates multiple 
muon tracks in an event by a single high energy muon 
which is adequate at very high energies 
for the variables used in this analysis. 
The single muon approximation predicts larger fluctuations in NPE
due to radiative energy losses of energetic muons, 
giving a rather conservative estimate
of the background passing rate. The two independent sets of
simulations with top-down and bottom-up approaches 
to describe the observational data complement each other 
improving the reliability of the background estimation.
%
\subsubsection{Background estimation with CORSIKA}
\label{subsec:corsika}
%

%
\begin{figure}[bt]
 \includegraphics[width=1.65in]{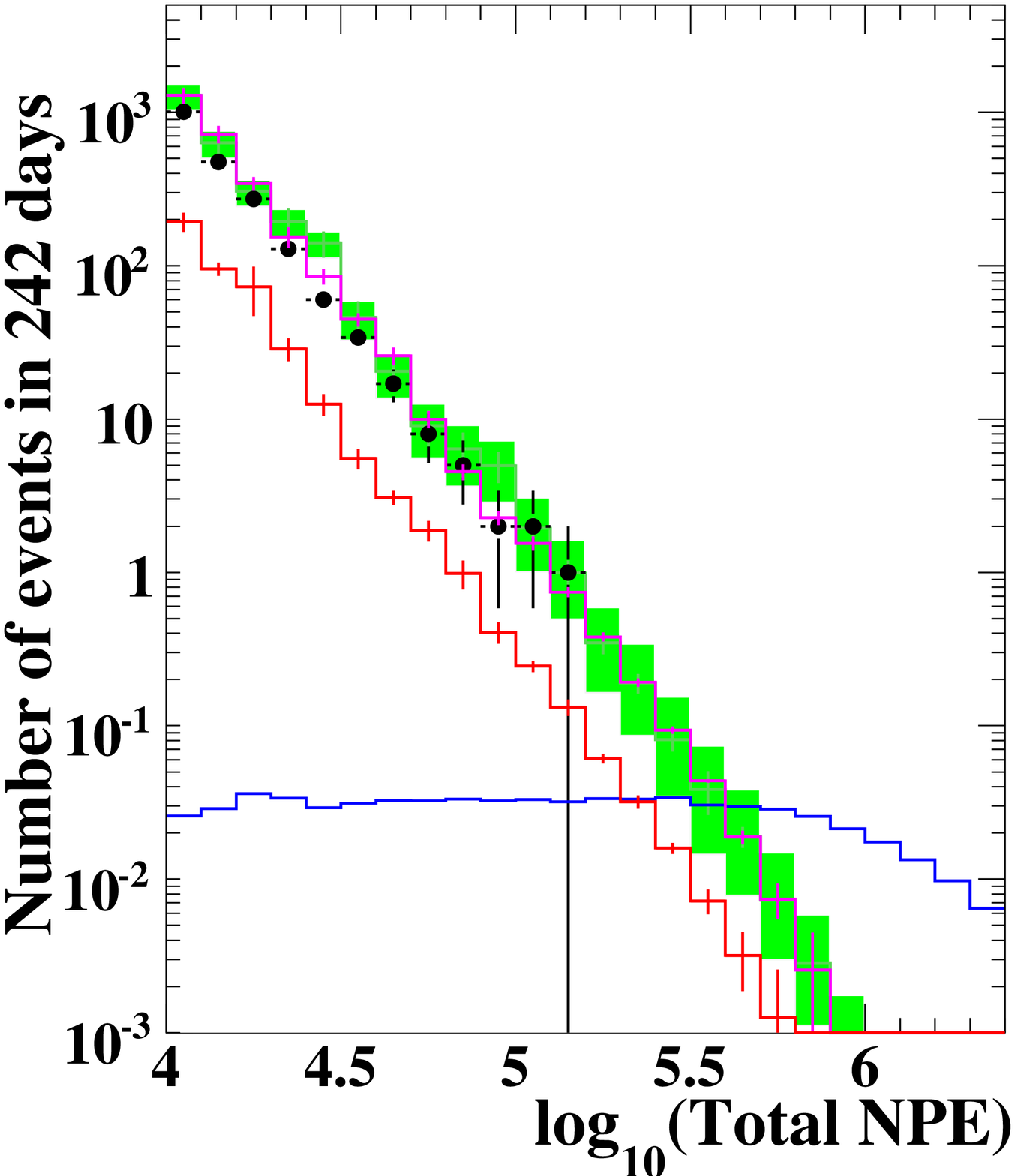}
 \includegraphics[width=1.65in]{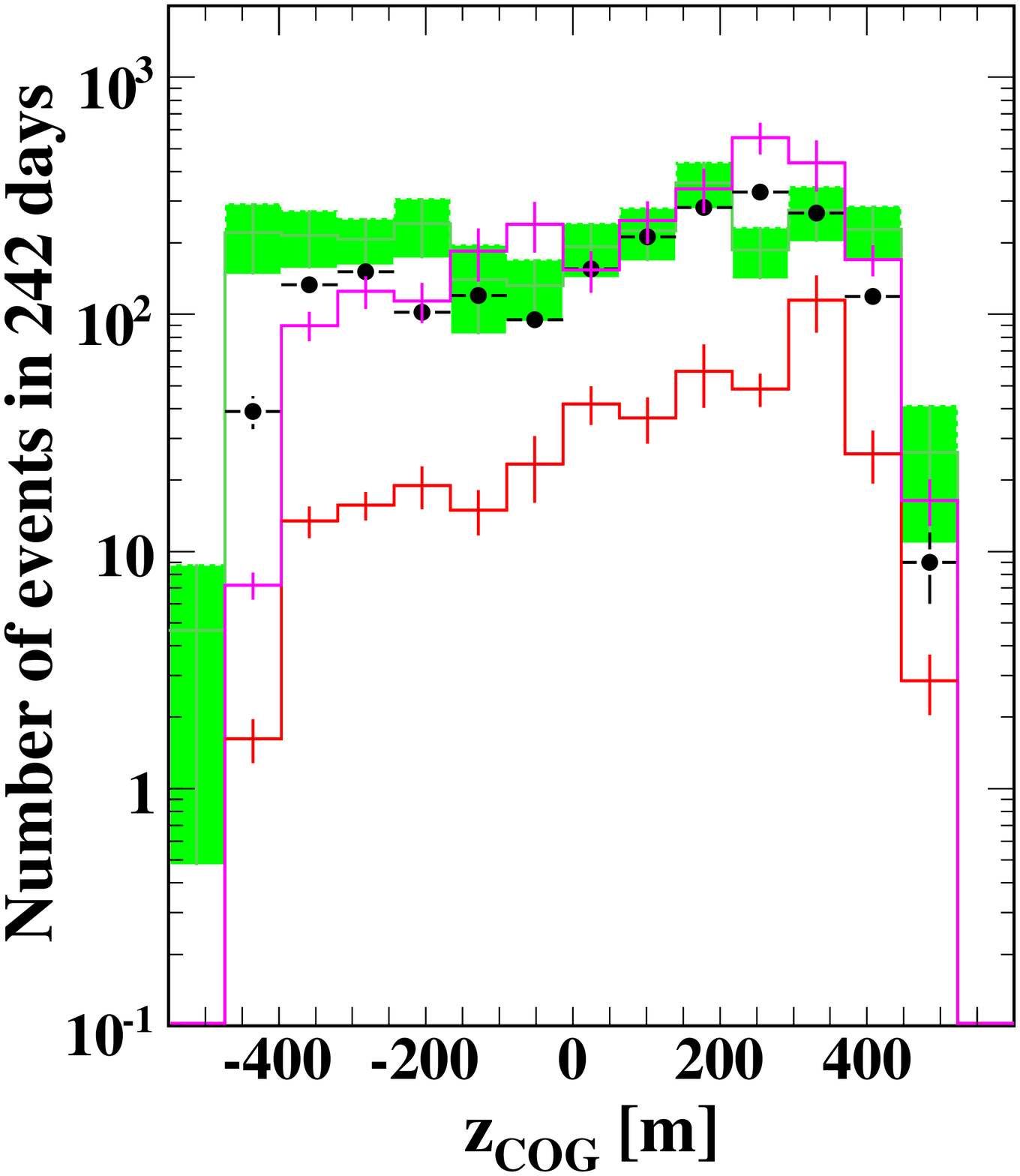}
 \caption{Event distributions from the level-3 
 samples as functions of NPE (left) and $z_{COG}$ (right). 
 The black dots represent observational data, 
 green boxes represent the empirical model 
 including uncertainty. 
 Red and magenta lines are CORSIKA samples with SIBYLL interaction model 
 and proton and iron primaries, respectively.
 The left panel also includes 
 the expected NPE distribution of events induced 
 by the cosmogenic neutrinos~\cite{yoshida93} shown by
 the blue line for reference.
 \label{Fig:Level3_BG_Comp}}
\end{figure}

Fig.~\ref{BGcompCorsika} shows distributions for data and simulations
at level-2 for NPE, the reconstructed zenith angle ($\theta$), and the 
center-of-gravity depth of the events ($z_{COG}$).
The CORSIKA NPE distribution is extrapolated to the higher NPE region.
Extrapolation was necessary mainly because of a
lack of simulated CORSIKA events at primary cosmic ray energies above $10^{10}$ GeV.   
The extrapolation accounts for the observed GZK cutoff 
at an energy around $5\times10^{19}$eV.
The NPE-weighted LineFit algorithm was used to reconstruct
zenith angle in this initial study.
The NPE-weighted LineFit is a simple minimization of 
$\chi^2 = \Sigma_i NPE_i (\vec{r_i}-\vec{r}_{COG}-t_i\vec{v})^2$,
where $t_i$ and $NPE_i$ represent, respectively,
the time of the first photoelectron and the number of photoelectrons
recorded by the {\it i}'th DOM at the position $\vec{r_i}$
and 
$\vec{r}_{COG} \equiv
(\frac{\Sigma_i NPE_i \;{x_i}}{\Sigma_iNPE_i}, \frac{\Sigma_i NPE_i
\;{y_i}}{\Sigma_iNPE_i}, \frac{\Sigma_i NPE_i \;{z_i}}{\Sigma_iNPE_i})$ 
is the 
NPE-weighted position of the center-of-gravity of the hits.
%
The fit ignores the geometry of the Cherenkov cone and the optical
properties of the medium and assumes light traveling with a velocity
$\vec{v}$ along a 1-dimensional path through the detector,
passing through the center-of-gravity.

The measured event rates are close to the 
simulated rates based on CORSIKA/SIBYLL with iron primaries and above 
those based on CORSIKA/SIBYLL proton data in most regions. 
A significant discrepancy can be found in the rate of events with 
cos $\theta$ $\leq 0.3$, {\it i.e.} events reconstructed
as horizontal or up-going, 
which is largely underestimated. 
Replacing SIBYLL with other hadronization models (e.g. QGSJET-II) 
does not change this behavior. 
The discrepancy may be due to 
a combination of uncertainties in the hadronic interaction models, 
cosmic-ray flux, and 
Cherenkov photon propagation in the glacial ice.
Since the horizon is the key region for the EHE neutrino search, 
the background estimations were supplemented 
by an empirical model fit to a subsample of the data.
%
\subsubsection{Construction of an empirical model}
\label{subsec:constElbert}
%
The empirical model is optimized to match the level-2 experimental data 
 ($10^{4} <$ NPE $< 10^{5}$). The possible signal region (NPE $\geq 10^5$) 
is not used to avoid bias. The model provides a relation between the 
NPE of an event and the cosmic-ray primary energy. 
Its convolution with the cosmic ray flux then gives the event rate
with a given NPE.
The cosmic-ray flux used in the present analysis is taken from the 
compilation in Ref.~\cite{nagano01}.

The model is based on the so-called Elbert formula~\cite{gaisser90} 
which parametrizes the mean multiplicity of muons with energies above a 
certain threshold $E_{\mu}$:
%
\begin{eqnarray}
N_{\mu} &=& {E_T\over E_0}{A^2\over \cos\theta'}\left({AE_{\mu}\over
 E_0}\right)^{-\alpha}\left(1-{AE_{\mu}\over E_0}\right)^{\beta},\\
E_T &=& 14.5\quad {\rm GeV} \nonumber
\end{eqnarray}
%
where $A$, $E_0$, and $\theta'$ are the mass number, the energy, and the zenith 
angle of the primary cosmic-ray~\cite{elbert79}.
The energy weighted integration of the formula relates the total energy
carried by a muon bundle $E_{\mu}^B$ to the primary cosmic-ray
energy $E_0$,
%
\begin{eqnarray}
E_{\mu}^B & \equiv & \int\limits_{\epsilon}^{E_0/A}{dN_{\mu}\over
 dE_\mu} E_\mu dE_\mu \nonumber\\
& \simeq & E_T {A\over \cos\theta'} {\alpha\over \alpha -1}
\left({A\epsilon \over E_0}\right)^{-\alpha +1},
\label{eq:Elbert}
\end{eqnarray}
%
assuming $AE_{\mu}/E_0 \ll 1$. 
%
Here, $\epsilon$ is empirically determined by fit to the observed data.
Assuming its corresponding energy at the IceCube depth,
$\epsilon^{\bf in-ice}$, is independent of zenith angle,
$\epsilon$ (and thereby $E_{\mu}^B$) can be calculated 
as a function of zenith angle
by taking into account the energy loss
during propagation through the Earth.
The optimization of the two parameters $\alpha$ and $\epsilon^{\bf in-ice}$ is 
performed by comparing the observed data to simulation 
of a single high energy muon with energy of $E_{\mu}^B$
in the NPE and zenith angle space, independently. $A=1$ is assumed
in the optimization.
The event distributions derived from the empirical model with optimized parameters 
($\alpha$ = 1.97 and $\epsilon^{\bf in-ice}$= 1500 GeV) are
given in Fig.~\ref{BGcompCorsika}. 
The green shaded region in the plot is obtained by allowing the model 
parameters to vary within $\pm$ 1 
$\sigma$ from their optimized values.
The discrepancies of $z_{COG}$ at large depths for the empirical model 
and at small depths for CORSIKA/iron seem to be due to vertical, 
down-going events because a restriction of the zenith angle, 
cos $\theta$ $<$ 0.8, improves the agreement in both cases. 
Since the majority of the EHE neutrino induced events is 
close to the horizon we can discard 
all events with cos $\theta$ $<$ 0.8 
without significant loss of signal efficiency (level-3 cut). 
The resulting distributions are shown in Fig.~\ref{Fig:Level3_BG_Comp}.

\begin{table*}[t]
  \caption{Number of events at analysis filter levels
    for 242.1 days in 2007. The simulation predictions for the atmospheric muon
    background using the CORSIKA-SIBYLL package, the empirical model, and
    that for the GZK cosmogenic neutrino model are also listed for comparison.
    Errors shown here are statistical only. See Sec.~\ref{sec:BG} and
 \ref{sec:search} for details.}
  \label{tab:analysis_filter}
  \centering
  \begin{tabular}{lccccc}
    \hline\hline
    Analysis filter levels & observational  data & empirical model & CORSIKA (iron) & CORSIKA (proton) & signal (GZK1~\cite{yoshida93}) \\ \hline
    level 3 ($\cos(\theta) < 0.8$) & 2014 & (2.65 $\pm$ 0.21)$\times 10^3$ & (2.68 $\pm$  0.19)$\times 10^3$ & (4.16 $\pm$ 0.40)$\times 10^2$ & (620 $\pm$ 7.3)$\times 10^{-3}$\\
    level 4 (EHE $\nu$ search) & 0 & (6.32 $\pm$ 1.37)$\times 10^{-4}$ &   (4.18 $\pm$ 1.29)$\times 10^{-4}$ & (1.44$\pm$ 0.58)$\times 10^{-4}$ &  (155 $\pm$ 1.4)$\times 10^{-3}$  \\  
   \hline\hline
  \end{tabular}
\end{table*}

%
%
%
\subsection{Search for EHE cosmogenic neutrino signal}
\label{sec:search}

\begin{figure*}
  \includegraphics[width=1.6in]{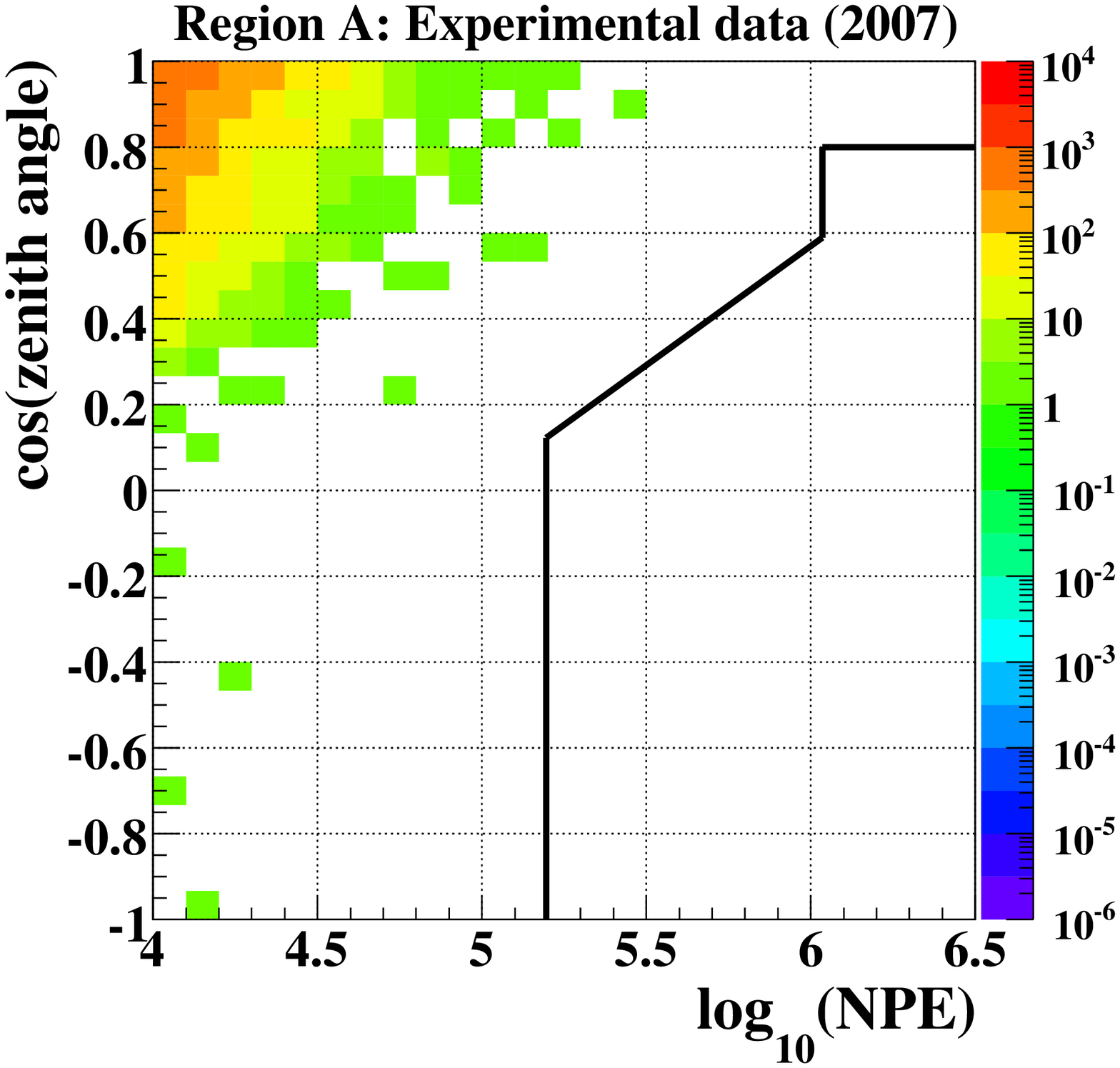}
  \includegraphics[width=1.6in]{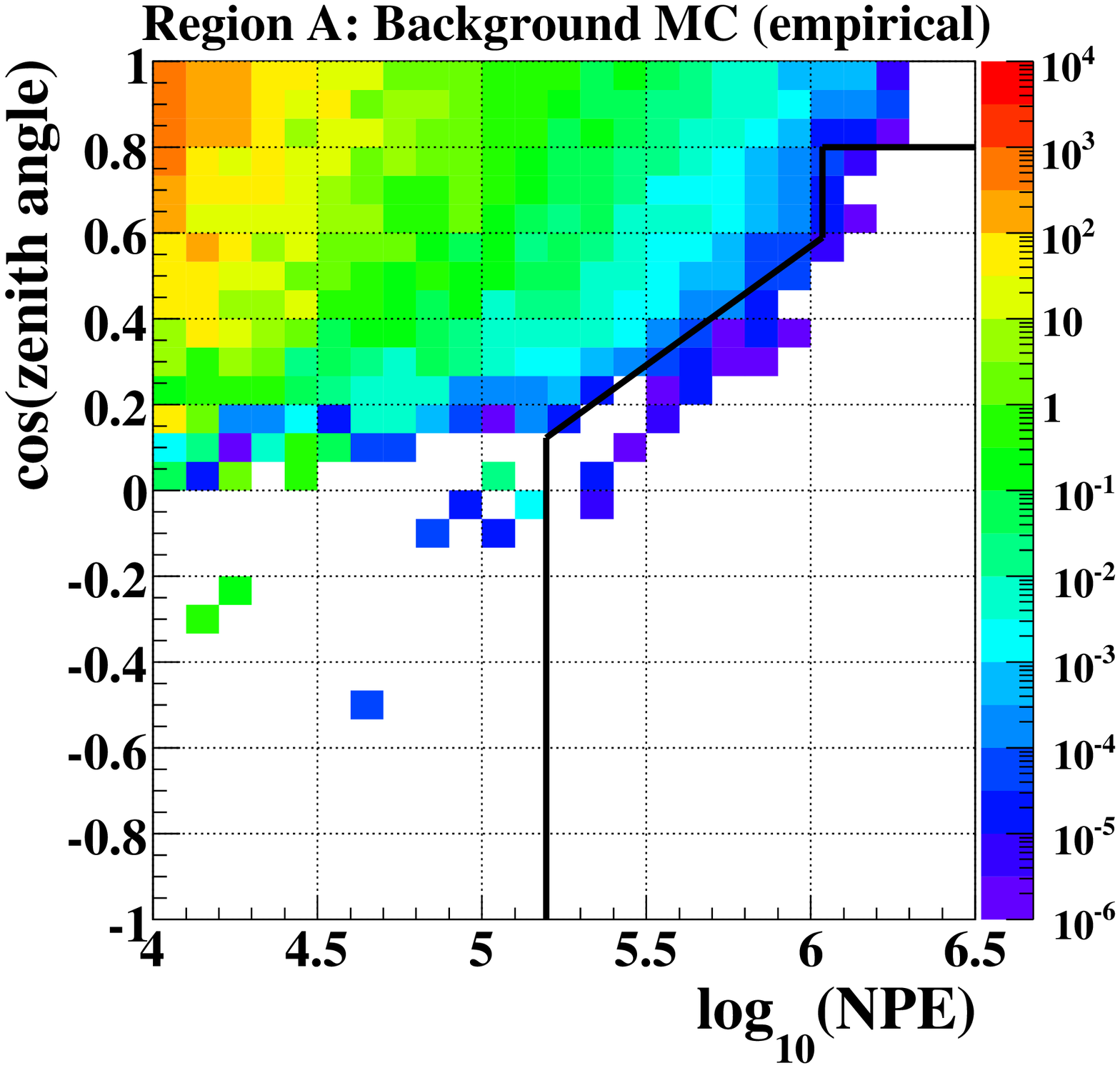}
  \includegraphics[width=1.6in]{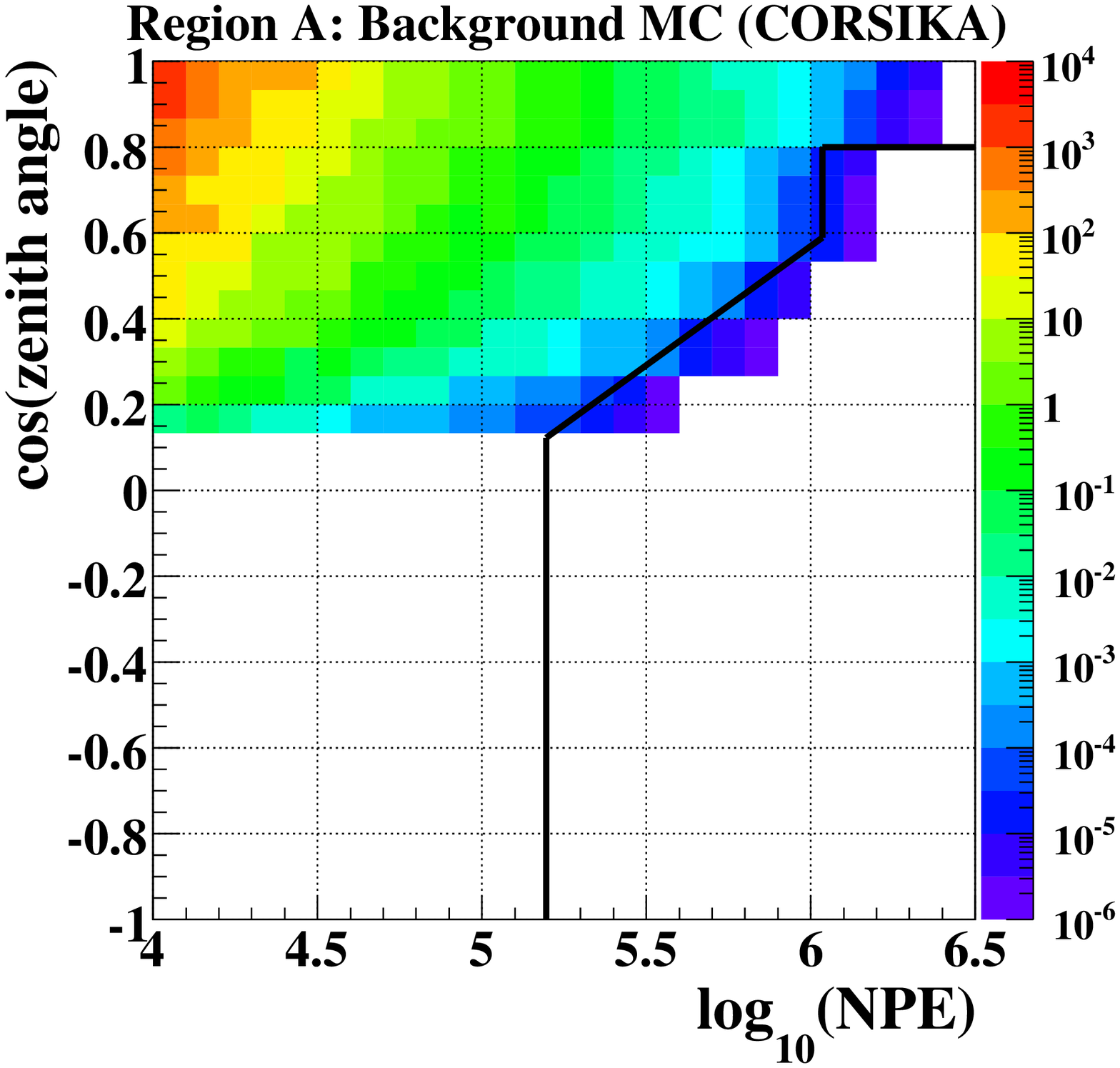}
  \includegraphics[width=1.6in]{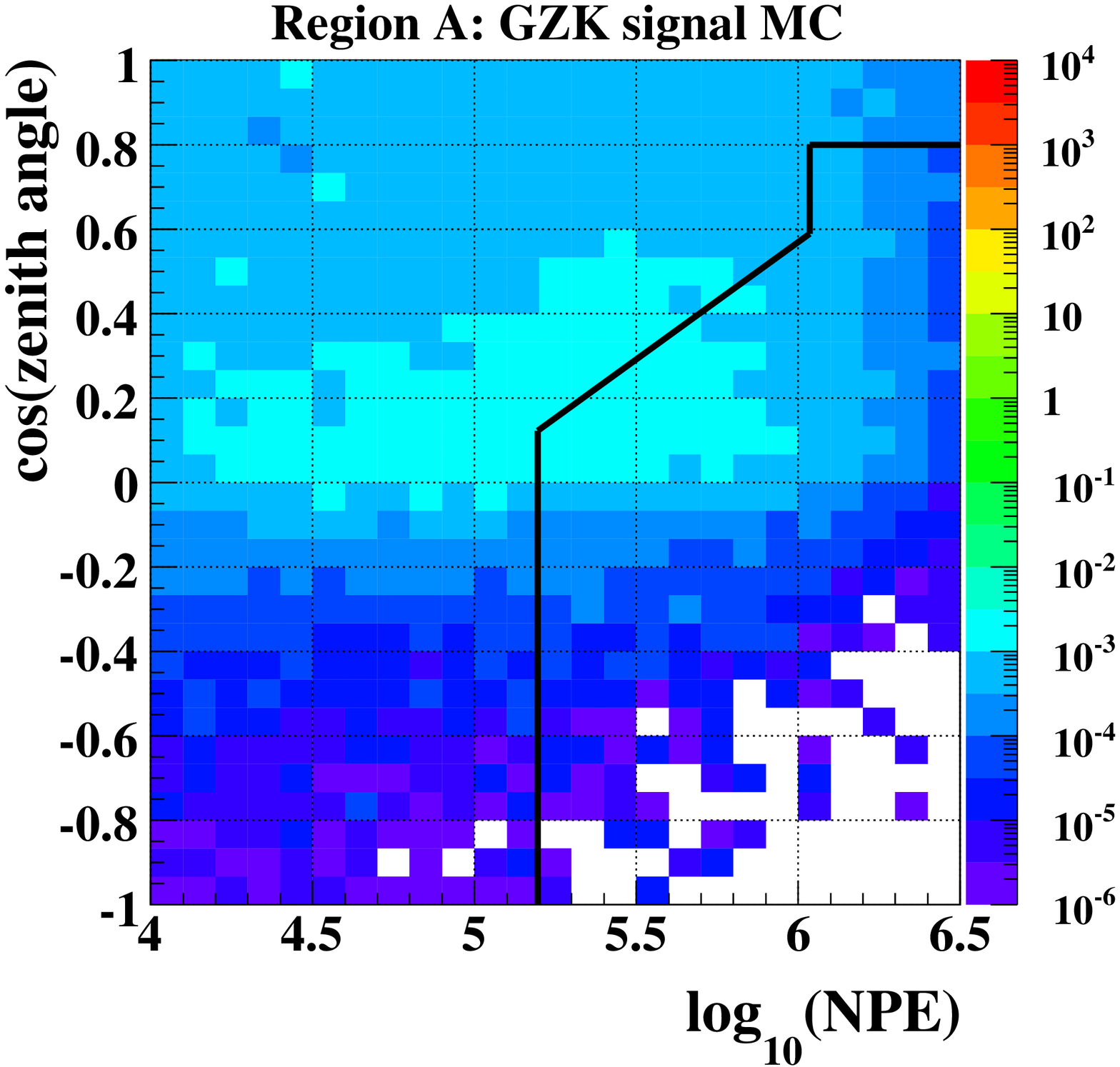}
  \includegraphics[width=1.6in]{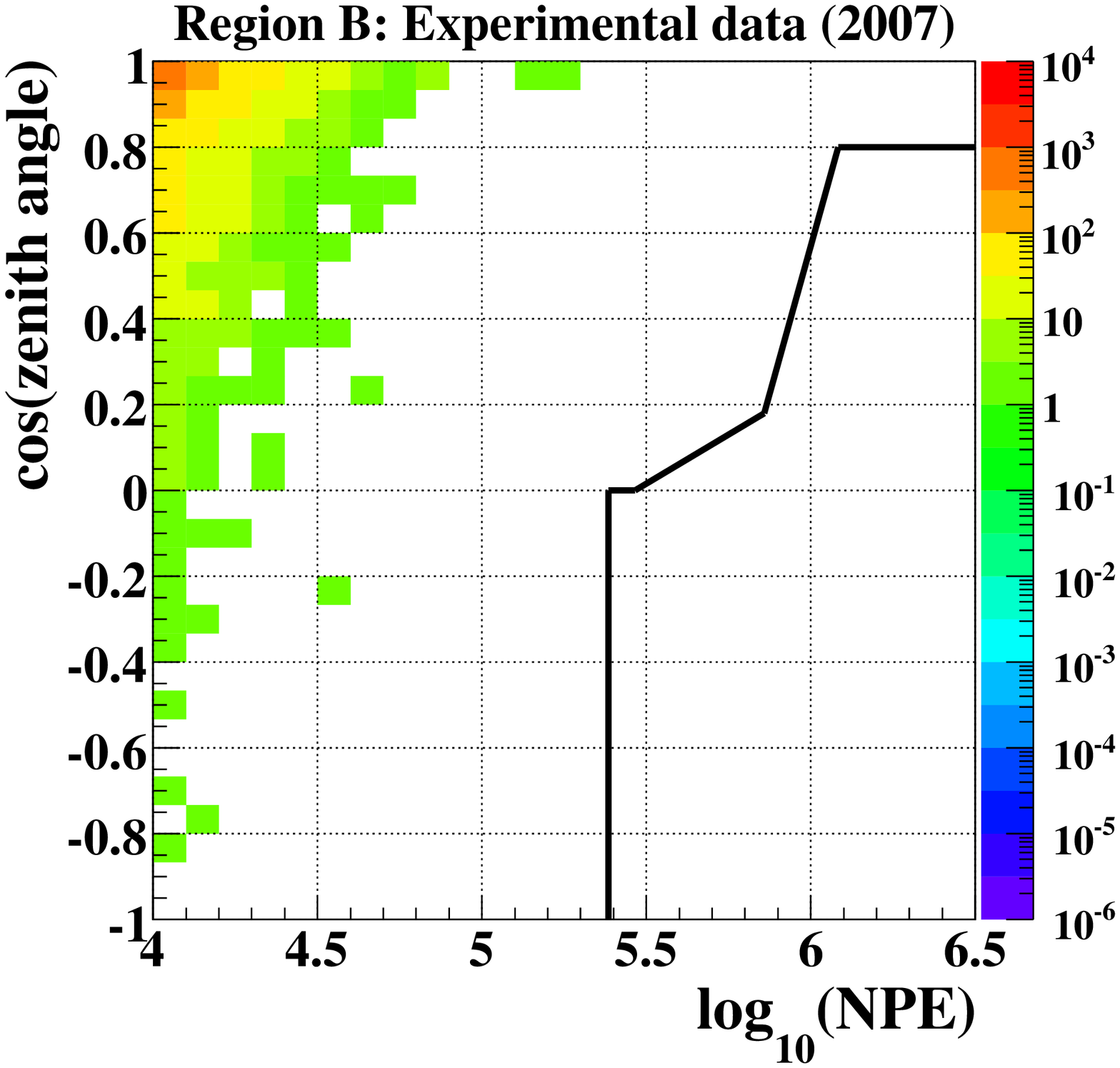}
  \includegraphics[width=1.6in]{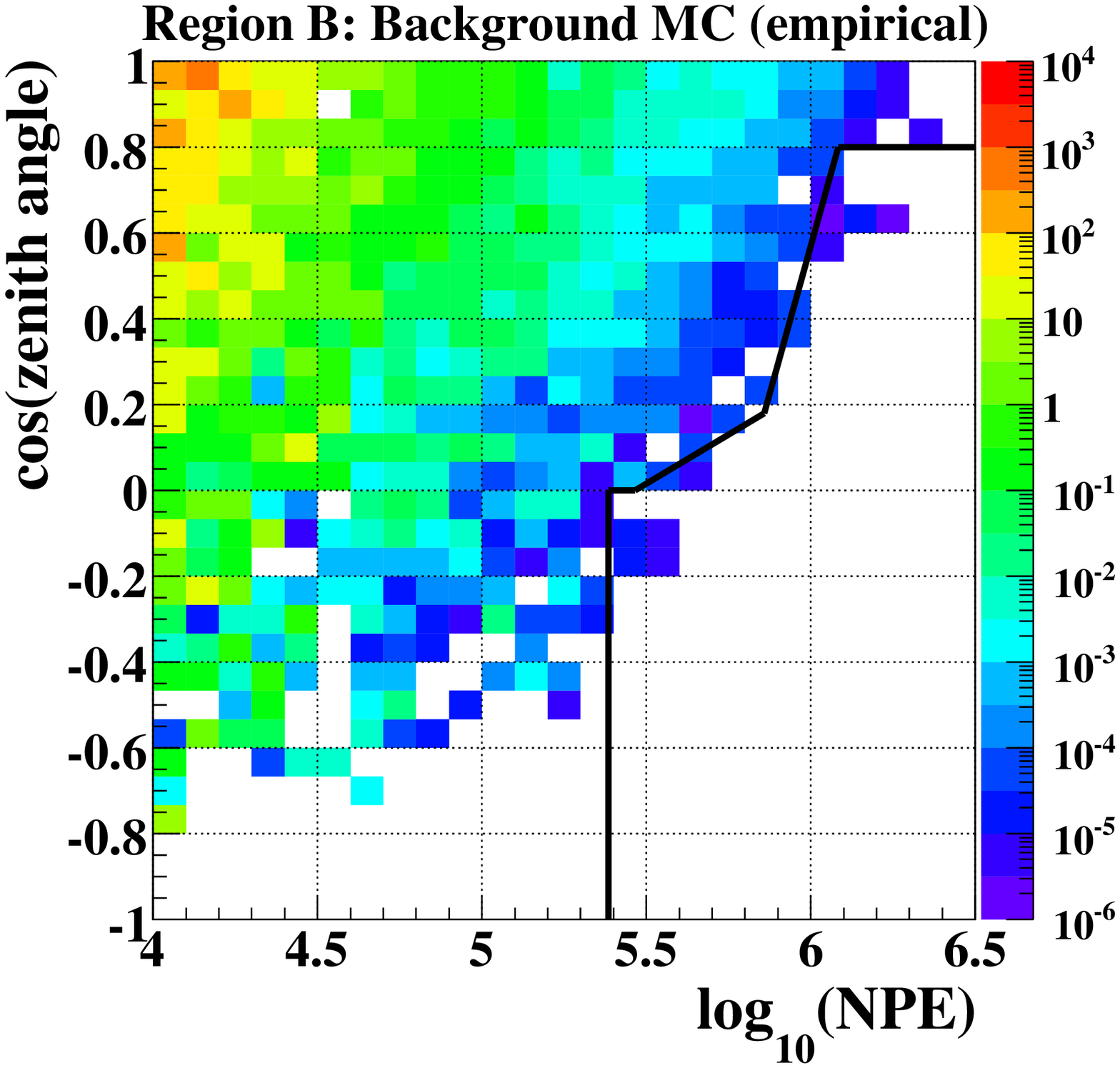}
  \includegraphics[width=1.6in]{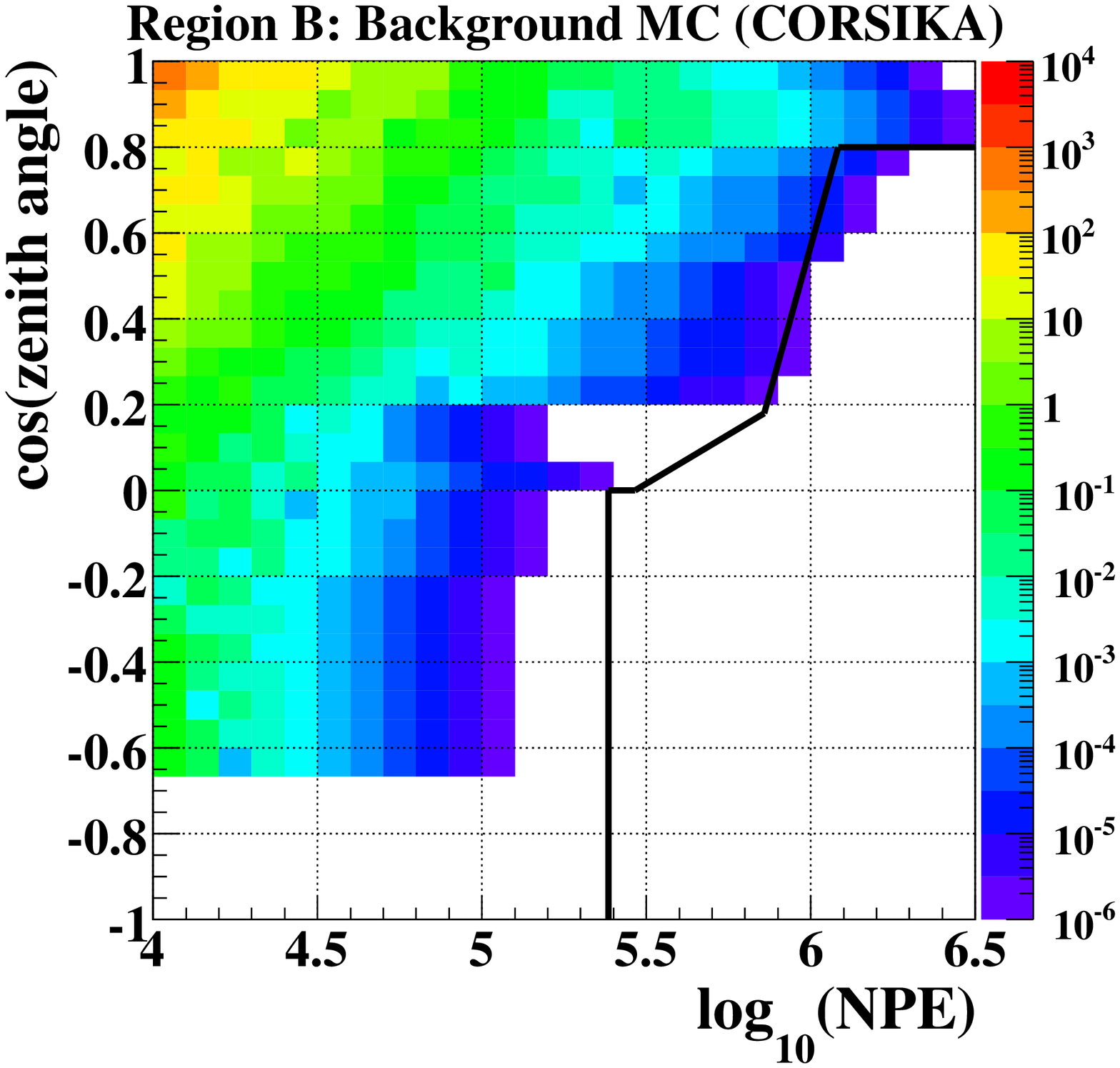}
  \includegraphics[width=1.6in]{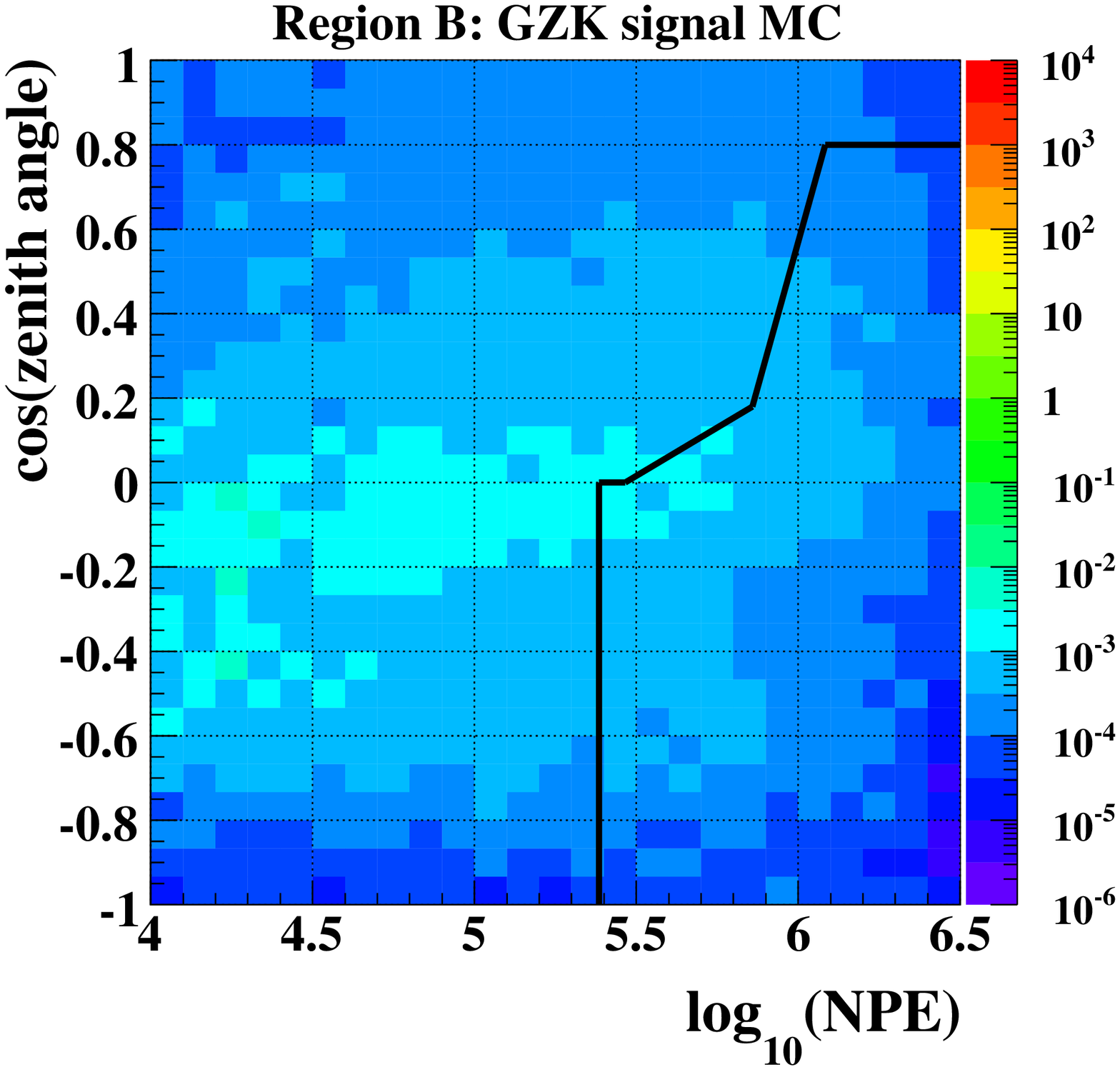}
  \caption{Event number distributions 
passing the level-2 selection cut (NPE$> 10^4$)
of the experimental data (left),
the background from the empirical model (middle left), the background from
CORSIKA SIBYLL with iron primaries (middle right), and 
the signal (right) on the NZ plane at the IceCube depth.
The upper (lower) panels show the distributions in the region~A (B).
The GZK neutrino flux~\cite{yoshida93} determines the event intensity
in the signal MC plot, adding all three flavors of neutrinos.
The series of thick lines in each panel indicates the level-3 ($\cos(\theta)<0.8$)
and the final level-4 cuts.
\label{Fig:NPEVsCosTheta}}
\end{figure*}

\begin{table}[bt]
 \caption{Expected event numbers passing the final level-4 selection criteria
 in the 2007 \mbox{IC-22} observation. Models include the GZK
 models~\cite{yoshida93,kalashev02,ESS} and the Z-burst model~\cite{yoshida98}.
 The predictions are normalized to a livetime of 242.1 days. Signal event
 numbers represent the sum over all three neutrino flavors.
 The first uncertainty is the statistical uncertainty determined
 by signal simulation statistics, and the second is
 the total systematic uncertainty from sources discussed in Sec.~\ref{sec:sys}.
  }
  \label{tb:Expect}
  \centering
  \begin{tabular}{lc}
    \hline\hline
    Models & Number of Events per 242.1 days \\
    \hline
    GZK1~\cite{yoshida93}  & (155 $\pm$ 1.4  $^{+24}_{-40}$ )$\times 10^{-3}$ \\
    GZK2~\cite{kalashev02}  & (248 $\pm$ 2.3  $^{+39}_{-65}$ )$\times 10^{-3}$ \\
    GZK3~\cite{ESS}  & (83 $\pm$ 0.8  $^{+13}_{-21}$ )$\times 10^{-3}$ \\
    Z-burst~\cite{yoshida98}  & (398 $\pm$ 3.4  $^{+63}_{-95}$ )$\times 10^{-3}$ \\
    \hline\hline
  \end{tabular}
\end{table}

The level-4 cut to eliminate the muon background is carried out in the
NPE-cos $\theta$ (NZ) plane.  
In accordance with the requirements of blindness, the cuts are finalized on simulated 
events alone without referring to real data. 
Because the optical properties of the glacial ice 
vary significantly with depth \cite{ice}, and because
the changing absorption and scattering lengths affect 
what IceCube observes, the final cuts are chosen to be depth dependent.  
The cuts are chosen based on the depth of 
the weighted center of gravity of the event,
$z_{COG}$.  
The distribution of events
in the NZ plane depends on $z_{COG}$. 
We divide the events into two groups according to their $z_{COG}$ as follows:

\begin{center}
  \begin{tabular}{c c}
    region A: & $-$250 $<$ $z_{COG}$ $<$ $-$50~m and $z_{COG}$ $>$ 50~m  \\
    region B: & $z_{COG}$ $<$ $-$250~m and $-$50 $<$ $z_{COG}$ $<$ 50~m  \\
  \end{tabular}
\end{center}

As seen in Fig.~\ref{Fig:NPEVsCosTheta}, 
region B contains a large number of horizontal and up-going mis-reconstructed 
background events, whereas the fraction of such events in region A is very small. 
Fig.~\ref{Fig:NPEVsCosTheta} also shows the distributions of the experimental data
and the simulated GZK neutrino induced signal events. 
The latter clearly accumulate near the horizontal direction 
regardless of the $z_{COG}$ position and have on average larger NPE 
than the background sample. 
The selection criteria to separate signal from background are 
determined for region A and B separately: 
For each bin of cos $\theta$ (width 0.1), a threshold NPE is set such that
the number of background events above the threshold is less than $10^{-4}$. 
Tighter cuts to further reduce the background would 
also reduce the signal 
to an undesirable degree.
The NPE thresholds in all the zenith angle bins
are then connected to each other to form a series of lines on the NZ plane, 
defining the final level-4 cut, as drawn in Fig.~\ref{Fig:NPEVsCosTheta}. 
The cuts were optimized using the empirical model for
background simulations, so we used extrapolated CORSIKA/iron data as a check
of the final background level. 
Figure~\ref{Fig:NPEVsCosTheta} also shows the distribution of background events 
in the NZ plane from the extrapolation.
Table~\ref{tab:analysis_filter} summarizes the number of events 
remaining in the analysis after each of the cut levels.

\begin{figure}
\begin{tabular}{cc}
  \includegraphics[height=1.77in,width=1.55in]{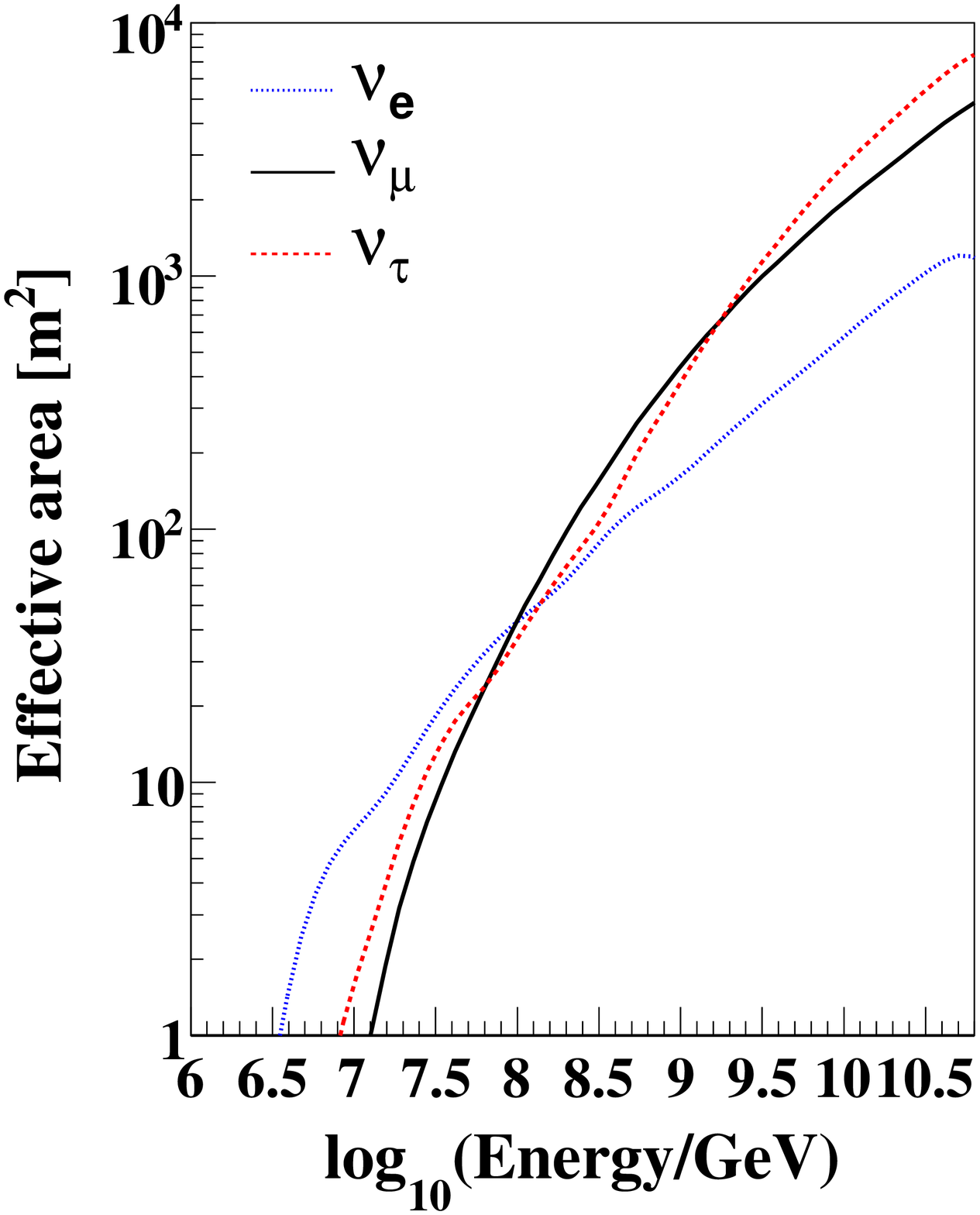}& \
  \includegraphics[height=1.77in,width=1.55in]{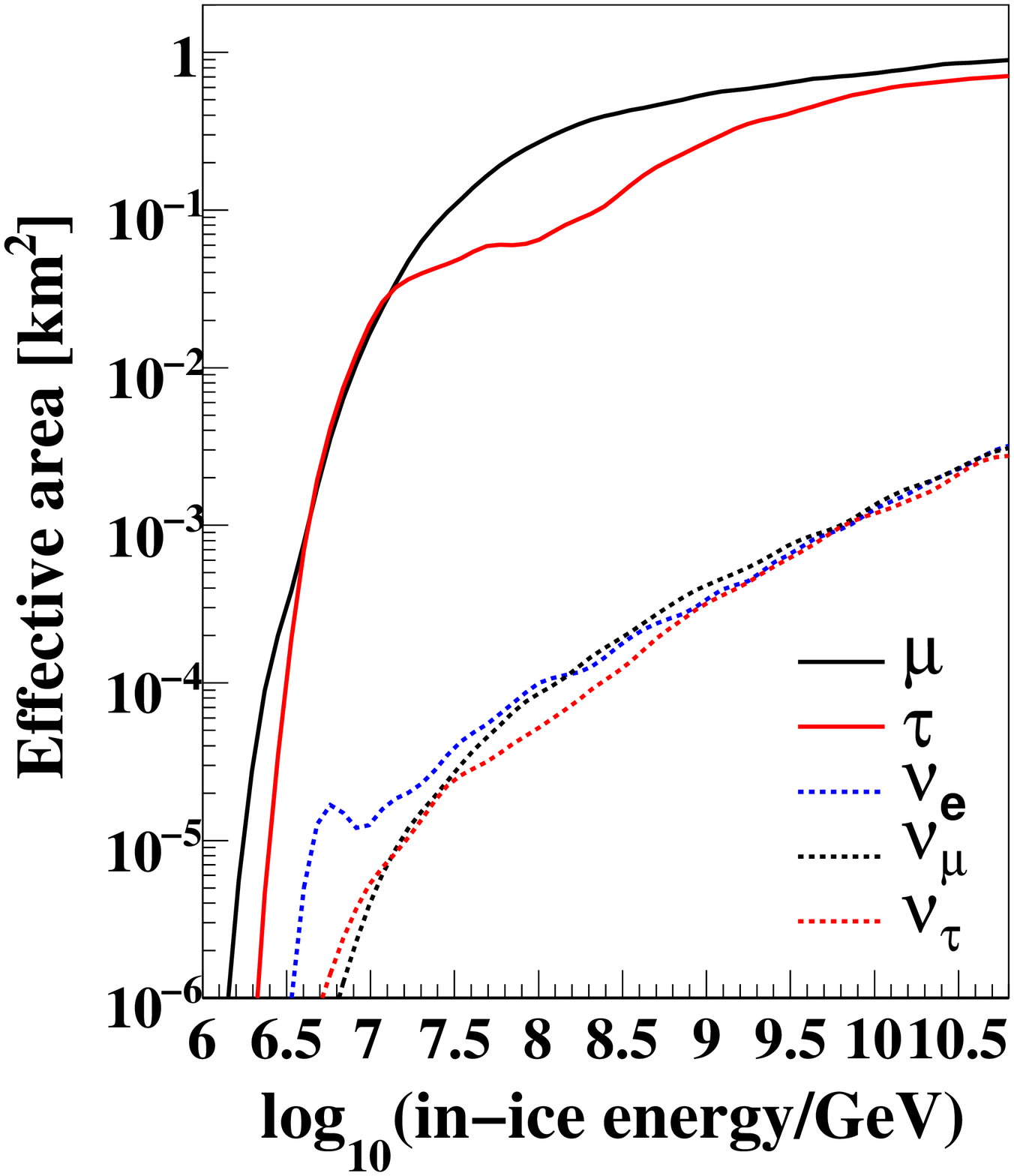} \\
\end{tabular}
  \vspace{-2mm}
  \caption{The effective area of \mbox{IC-22} for EHE neutrino search.
    The left panel shows the 4$\pi$ solid angle averaged
    area as a function of neutrino energy at the Earth surface.
    The right panel shows the corresponding effective area 
    for particles at 880 m from the IceCube center
    entering into the \mbox{IC-22} fiducial volume. Muons and taus in this plot
    are secondary particles produced by neutrinos before reaching the neighborhood
    of the detector array. The energy here are defined as \mbox{in-ice energy}.
}
    \label{effArea}
\end{figure}

The effective area as a function of energy at the Earth surface 
for each neutrino flavor is shown in 
the left panel of Fig.~\ref{effArea}, averaged 
over all solid angles.  The area increases with the energy owing to
the increasing neutrino interaction cross sections and the increased
probability of observing the interactions, which is different
for each flavor.

At low energies, most of the $\nu_\tau$ signal comes from events where the
$\nu_\tau$ interacts in the detector, or the $\tau$ decays in it.  
At higher energies, $\tau$
energy loss becomes large enough that through-going taus also pass the cuts.
Contributions from $\nu_{\mu}$ and $\nu_{\tau}$
dominate over $\nu_e$ in the energy range above $\sim 10^8$ GeV, as the
secondary produced muons and taus can travel long distances to reach the
detection volume. This trend is reversed at lower energy 
where tau and muon energy losses are smaller, and 
$\nu_e$'s can deposit all of their energy into the detector volume.
The effective
area for $\nu_{\tau}$ is larger than that for $\nu_{\mu}$ at low
energies, because of the events where taus decay inside the
detector.
At the highest energies,
because of the larger mass of taus and 
increase of the tau decay time with energy,
the tau range is longer than that of muons, 
leading to a larger effective area.
%

The right panel in Fig.~\ref{effArea} shows the effective area as a
function of the in-ice energy
(''in-ice area''). 
It represents the probability of detection of incoming particles
with the present analysis.
The area for incident muons and taus gradually increases with
energy but is limited essentially by the physical cross section of 
the IC-22 array, $\sim$ 0.5 km$^2$. 
Because the Cherenkov yield of taus is smaller than muons with the same energy
due to the smaller radiative energy loss, the detection probability
of incident taus is lower, leading to the smaller in-ice area.
Incoming neutrinos must interact to yield Cherenkov light to be detected.
Therefore the neutrino effective area becomes 
much smaller than that for muons or taus.

The expected number of signal events for various neutrino production models 
after the level-4 cut are summarized in TABLE~\ref{tb:Expect}. 
GZK1~\cite{yoshida93} represents 
the case of a moderately strong source evolution, $(z+1)^m$ with $m=4$ 
extending to $z=4.0$, while GZK2~\cite{kalashev02} assumes $m=5$ up to $z=2.0$, 
and GZK3~\cite{ESS} uses $m=3$ with a slightly different parametrization 
and a cut-off structure. 

%
%
\section{The Systematics}   
\label{sec:sys}

This search is based on the event-wise NPE and
reconstructed zenith angle. The main systematic uncertainties derive
from 1) the necessity to extrapolate the empirical fit to data by
approximately an order of magnitude in NPE to estimate the background
rate at the highest energies and from 2) the uncertainty of the absolute NPE scale.
Table~\ref{tb:sys} lists the sources of statistical and systematics errors.
\begin{table}
  \caption{List of the statistical and systematic errors. The signal rate
  is estimated by assuming the high evolution flux (m,Z$_{{\rm max}}$)$ = (4,4)$
  in Ref~\cite{yoshida93}.}
  \label{tb:sys}
  \begin{tabular}{ccc}
    \hline\hline
    Error source & background & signal (GZK) \\
                 & rate & rate \\
    \hline
    Statistical error  & $\pm$22\% & $\pm$0.9\% \\
    Detector sensitivity  & - &  $\pm$8\% \\
    Yearly variation  & $\pm$17\% &  - \\
    Empirical model  & +99/-59\% &  - \\
    Background model dep.  & $\pm$15 \% & -  \\
    NPE yield & -  & +0/-21\% \\ 
    Neutrino cross section  & - &  $\pm$9\% \\
    Photo-nuclear interaction & - &  +10\% \\
    LPM effect  & - &  $\pm$1\% \\
    \hline \hline
    Total  & $\pm$22\% (stat.) & $\pm$0.9\% (stat.)  \\
    & +102/-63\% (sys.) & +16/-26\% (sys.) \\
    \hline\hline
  \end{tabular}
\end{table}

\subsection{Uncertainties in the background rate estimation}
\label{subsec:bg-uncertainty}
The largest uncertainty in the background rate estimate arises from the
fact that the parameters of the empirical model were optimized for the
observed events with $10^{4} <$ NPE $< 10^{5}$ after level 2
selection. The limited statistics of this sample results in
uncertainties on the parameters. The model was then extrapolated to a
higher NPE region for the determination of the level-4 cut.

Allowing the parameters to vary within $\pm$ 1 $\sigma$ changes the
background rate by between  $-$59\% (for the softest possible
NPE spectrum after the level-4 cuts) and $+$ 99\% (for the hardest possible
NPE spectrum). Uncertainties in the detector sensitivity are
incorporated by the parametrization. The difference in the
background level estimated with the extrapolated CORSIKA/iron and the
empirical model can be taken to indicate the level of systematic
uncertainty due to model dependence. This uncertainty is
approximately $\pm$15\% and can be assumed to 
include the possible contribution from charm decay.
An uncertainty associated with the high
energy hadronic interaction model is evaluated using simulated muon
bundle intensity from SIBYLL and QGSJET II with iron primaries and found
to be $\pm$ 4\%, which is negligible. 
An additional uncertainty of $\sim$17\% arises 
from the seasonal variation of the atmospheric muon rate
as the signal selection criteria 
are based upon the season-averaged 
data.

\subsection{Uncertainties of the signal rate estimate}
\label{subsec:NPE-uncertainty}
The uncertainty in the relationship 
between measured NPE and the energies of
charged particles 
is the largest systematic error 
affecting the signal event rate.
It 
is the consequence of our limited understanding of the detector sensitivity,
the photon propagation in ice, and the detector response to bright
signals. It is evaluated using absolutely calibrated in situ light
sources and amounts to a possible overestimation of NPE in simulation by 18.5\%,
which leads to decrease of the GZK signal rate by $\sim$21\%. 

Uncertainties of the relevant particle interactions
in the EHE regime also add systematics in the signal rate
estimation. 
The expected event rate
scales nearly linearly with the neutrino-nucleon inelastic cross section
in the EHE range. This scaling has been confirmed by 
numerical studies, artificially increasing the cross section.
The cross section uncertainty has been recently reduced
to be around $\pm 9$\% with the inclusion 
of the most recent data from HERA and modern
parton distribution functions~\cite{amanda08}.
Another systematic error
arises from the photonuclear cross section of EHE muons and taus.
The present calculation used the model 
by Bugaev and Shlepin~\cite{BB}, rewritten in
Ref.~\cite{igor04}, that includes
a relatively reliable soft nonperturbative component
and a 
less certain hard perturbative part. Ignoring
the hard component in the simulation gives the most conservative
estimate of the uncertainty and leads to a 10\% event rate increase. 
The suppression of bremsstrahlung and pair production due to the LPM
effect~\cite{lpm}, for the relevant electron energies of $10^{9-10}$ GeV,
increases the effective radiation length of the electromagnetic cascade to
$O(30\sim100)$~m from $\sim$36~cm~\cite{kleinlpm}.
Because the value is still comparable to the IceCube DOM separation,
and the contribution from $\nu_e$ constitutes $\leq$20\% of the total event
rate in this energy range, the LPM effect has a negligible impact
on the event rate. This has been confirmed by a special simulation study
on $\nu_e$ including the LPM cascade elongation. 

%
%
\section{Results}
\label{sec:results}

\begin{figure}
  \centering
  \includegraphics[width=3.4in]{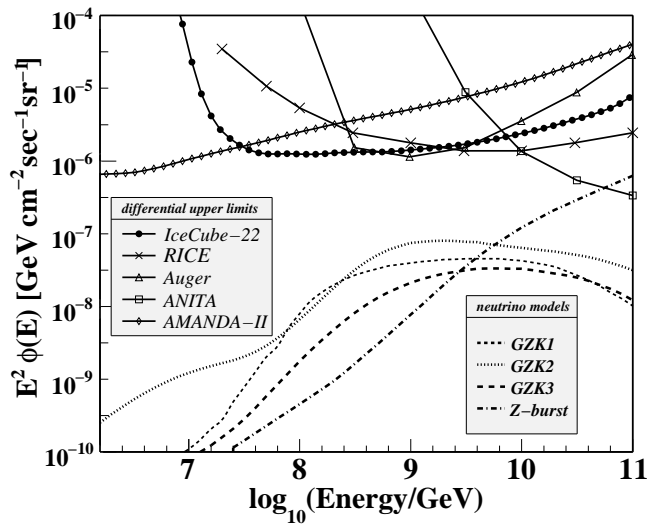}
  \caption{The all flavor neutrino flux differential limit 
    from the \mbox{IC-22} EHE analysis (filled circles). The systematic errors
    are included. Also the various model predictions are shown for comparison:
    GZK model~1~\cite{yoshida93} (short dashed line), GZK
 model~2~\cite{kalashev02} (dotted line), 
    GZK model~3~\cite{ESS} (long dashed line), Z-burst model~\cite{yoshida98}
 (dashed dot line).
    The model independent differential upper limits by other experiments are 
    also shown for Auger~\cite{auger} (open triangles), 
    RICE~\cite{rice} (crosses), ANITA~\cite{anita} (open squares),
    AMANDA~\cite{AMANDA} (rhombi).
    Limits from other experiments are converted to the all flavor limit
    assuming full mixing neutrino oscillations and 90\% C.L when necessary.
    \label{Fig:sensitivity}}
 \end{figure}

No events are observed in the final data sample taken in 2007 with
a livetime of 242.1 days when applying the final level-4 selection criteria,
which is consistent with the expected number of background events
of $6.3\times 10^{-4}$. 
We choose to present the resulting all flavor EHE neutrino upper limit in the 
quasi-differential form independent of the neutrino production model. 
Assuming full mixing due to oscillations, the experimental 90\% confidence level upper 
limit is obtained by setting 2.44 events~\cite{feldman98}
for an upper bound of the number of events observed with bin width of a decade of energy
with condition that energy dependence of neutrino flux 
multiplied by the effective area behaves 
as $\sim 1/E$~\cite{rice}.
This limit is presented in Fig.~\ref{Fig:sensitivity} 
including the systematic errors.
The plot indicates that the EHE neutrino search by the IceCube observatory
is most sensitive to the neutrinos with energies on
the Earth surface ranging between about $10^8$ and $10^9$ GeV.
The absence of signal events in the sample of 242.1 days of effective
livetime results in a 90\% C.L. differential 
upper limit on the neutrino flux of 
$E^2 \phi_{\nu_e+\nu_\mu+\nu_\tau}\simeq 1.4 \times 10^{-6}$
GeV cm$^{-2}$ sec$^{-1}$ sr$^{-1}$ on average for neutrinos with an energy of 
$3\times 10^7\leq E\leq 3\times 10^9$ GeV.
Here $\phi_{\nu_e+\nu_\mu+\nu_\tau}$ denotes the differential flux
of the sum over all three neutrino flavors, 
{\it i.e.} number of neutrinos per unit energy, area, time
and solid angle.

The quasi-differential limit in Fig.~\ref{Fig:sensitivity} 
takes into account the systematic 
uncertainties. 
The background rate stays negligible $O(10^{-3})$ 
even including the systematic uncertainty and the resultant
upper limit is unchanged. The signal rate uncertainty 
is strongly dominated by the uncertainty of the NPE yield which 
influences the number of expected signal events as a function of the 
neutrino energy.
The upper limit is calculated
by reducing NPE by 18.5\% 
in the signal simulation to account for
this factor. 
All the other sources of systematic error only slightly 
change the signal passing rate; they are
independent of energy.  They are included in the analysis by uniformly scaling the
effective area in the limit calculation.

The present limit is
approximately a factor
of 20-30 higher than the intensity range expected in the 
GZK cosmogenic neutrino production models~\cite{yoshida93,kalashev02,ESS}, 
as one can see in Fig.~\ref{Fig:sensitivity}.
The current limit for 242.1 days of observation is comparable 
to the Auger~\cite{auger} and HiRes~\cite{hires} bounds
by their multiple year operation.

%
%
\section{conclusions}  
\label{sec:summary}
The present work has demonstrated that the IceCube neutrino observatory
is capable of searching for signatures of EHE cosmogenic neutrinos with 
relatively straightforward event selection methods.
The 
model independent differential upper limit 
obtained with 242.1 days of observation in 2007, 
with approximately one quarter of the completed detector
is $E^2 \phi_{\nu_e+\nu_\mu+\nu_\tau}\simeq 1.4 \times 10^{-6}$
GeV cm$^{-2}$ sec$^{-1}$ sr$^{-1}$ for neutrinos with an energy of 
$3\times 10^7\leq E\leq 3\times 10^9$ GeV.
This is approximately
a factor of 20 higher than the
predicted GZK neutrino flux from relatively strongly evolved sources.
In the future, data taken by IceCube with 40 to 86 strings 
operating should lead to a detection of cosmogenic neutrinos 
or a greatly improved limit.

\begin{acknowledgments}
We acknowledge the support from the following agencies:
U.S. National Science Foundation-Office of Polar Programs,
U.S. National Science Foundation-Physics Division,
University of Wisconsin Alumni Research Foundation,
U.S. Department of Energy, and National Energy Research Scientific Computing Center,
the Louisiana Optical Network Initiative (LONI) grid computing resources;
Swedish Research Council,
Swedish Polar Research Secretariat,
Swedish National Infrastructure for Computing (SNIC),
and Knut and Alice Wallenberg Foundation, Sweden;
German Ministry for Education and Research (BMBF),
Deutsche Forschungsgemeinschaft (DFG),
Research Department of Plasmas with Complex Interactions (Bochum), Germany;
Fund for Scientific Research (FNRS-FWO),
FWO Odysseus programme,
Flanders Institute to encourage scientific and 
technological research in industry (IWT),
Belgian Federal Science Policy Office (Belspo);
University of Oxford, United Kingdom;
Marsden Fund, New Zealand;
Japan Society for Promotion of Science (JSPS);
the Swiss National Science Foundation (SNSF), Switzerland;
A.~Kappes and A.~Gro{\ss} acknowledge support by the EU Marie Curie OIF Program;
J.~P.~Rodrigues acknowledges support 
by the Capes Foundation, Ministry of Education of Brazil.
This analysis work has been particularly supported 
by the Japan-US Bilateral Joint Projects
in the Japan Society for the Promotion of Science.

\end{acknowledgments}

\end{document}